\documentclass[12pt,reprint,prb]{revtex4-1}
\usepackage{graphicx}%\input epsf
\usepackage{url}
\def\vecp#1{\vec {#1\mkern-1mu}\mkern 3mu{}'}
\def\subrm#1{{\hbox{\small #1}}}
\def\tinyrm#1{{\hbox{\tiny #1}}}
  % torque in Newtonian mechanics is 45.20.da

\begin{document}
\title{Torques without Rotation: the Right-Angle Lever}
\author{Joel A. Shapiro}
\affiliation{Department of Physics and Astronomy,\\
Rutgers University, Piscataway, NJ 08854}
\email{shapiro@physics.rutgers.edu}
\date{June 19, 2012}

\begin{abstract} 
An extended body subject to external forces which exert zero net force and
zero total torque in the rest frame, may experience a nonzero torque in 
another inertial frame, and nonetheless does not rotate. Long known as 
the Trouton-Noble or right-angle lever paradox, there has been extensive 
discussion and indeed controversy, but a clear understanding 
comes from a suitable treatment of angular momentum and simultaneity. 
\end{abstract} 
\maketitle

\section{Introduction}

Special relativity presents many situations with consequences that
seem paradoxical to someone accustomed to classical, Galilean
kinematics. Most often the relativity of simultaneity is the source of
the problem. Elementary courses discuss as examples the twin paradox,
the pole-vaulter entering the barn, and the sled falling through a
smaller hole in the ice. But the curious behavior of torques on a
rigid body, though discovered and explained in the first decade of
relativity, is not generally discussed in undergraduate courses.

The paradox arose from an attempt before Einstein's paper to find the
motion of the Earth through the aether by the torque aligning a
charged parallel plate capacitor perpendicular to the
motion\cite{TNexp}. Of course no effect was found. An equivalent
purely mechanical situation has a rod under compression or tension.

In exploring the angular momentum of a rigid body in elementary courses,
we claim that internal forces do not contribute to
the total torque and do not affect the angular momentum. This is because
we assume
action and reaction are equal and opposite, and {\em also act along a
  common line}, so as to have the same moment arm. This is called the 
{\em strong form} of Newton's third law. 

When two equal and opposite external forces act on different points of
a rigid body, such as two forces trying to stretch a thin rod, whether
they act along the same line or not depends on reference frame. If, in
the rest frame of the rod, they act along the direction of the rod,
they will produce no torque, and the momentum and angular momentum of
the rod will be conserved, in that frame. But transforming to a frame
with respect 
to which the rod is moving at an angle $\alpha$ relative to its length, 
the forces change direction and are generally not collinear, and
there is a torque. 

\section{The Right-Angle Lever}

An equivalent situation which is simpler to analyze, known
as the ``right-angle lever paradox'' was first discussed by
Lewis and Tolman in 1909\cite{LTprnnm,Tolman}.
Consider a rigid angle lever as shown, with 
%\begin{enumerate}
\begin{itemize}
\item a force $F_x$ acting at
the upper end A, at height $y$ above the origin at point O,
\item a force $F_y$ acting at the right end B, a distance $x$ from the
 origin, and 
\item a counterbalencing pair of forces $R_x$ and $R_y$.
\end{itemize}
%\end{enumerate}
\smallskip\noindent

\includegraphics[width=1.8in]{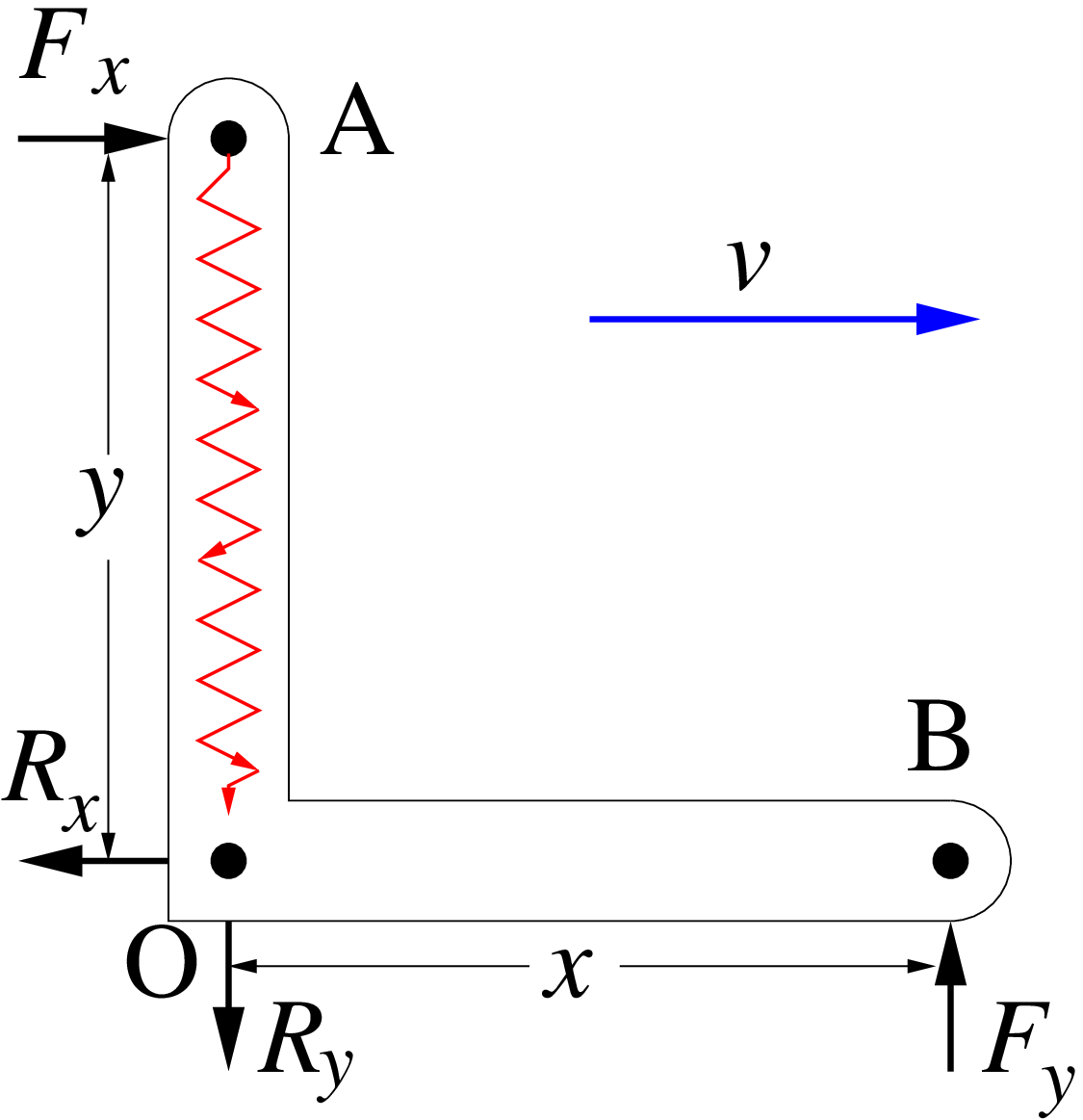}

In the rest frame of the
angle lever, let us take $x=y=L$, $F_x=F_y=R_x=R_y$ in magnitudes
(which we will call $F$), so that the
is no net force or net torque on the lever, and therefore the lever
remains at rest with zero momentum and zero angular momentum.

Now consider this from an observer ${\cal O}'$ for whom the lever
is moving right 
with velocity $v$. Distances perpendicular to the motion are unchanged, so
$y'=L$, but distances parallel are contracted, so $x'=L/\gamma$

How do forces transform\cite{Note1}? The 
Minkowski 4-force $f^\mu$ is a 4-vector,
but this is not the usual force. For a particle, the force $\vec F$ as
measured in some frame is $d\vec P/dt$ where $\vec P$ and $t$ are the 
particle's momentum and the frame's time, respectively, but the
4-force is $f^\mu=dP^\mu/d\tau$, where $d\tau = dt/\gamma$ is the
proper time for the particle on which the force acts. 
If we are considering a force acting on
one point of a rigid body,  we mean the proper time as measured
by that point of the body. So for the spatial components, 
$\vec f = \gamma \vec F$.

Now for our lever in its rest frame, $\gamma=1$, 
$f_A^x =F=f_B^y=-r^x=-r^y$, and all other components of the
4-forces are zero, including the zeroth components, which are zero
because the forces are doing no work. The Lorentz transformation to
the ${\cal O}'$ frame is 
$$f^{\prime\,0} = \beta\gamma f^1,\quad 
f^{\prime\,1} = \gamma f^1,\quad
f^{\prime\,2} =  f^2,\quad
f^{\prime\,3} =  f^3 = 0,$$
so the 4-force at A is $f'_A = (\beta\gamma F, \gamma F,0,0)$,
and $r'_x = -f'_A$, while the 4-force at the right is 
$f'_B = (0,0,F,0)$. As the 3-force is $1/\gamma$ times the 4-force
spatial components, we have $F'_{A\,x}=F=-R'_x$, $F'_{B\,y} = F/\gamma=-R'_y$,
and the total clockwise torque is $y'F'_{A\,x}-x'F'_{B\,y} =
LF(1-\gamma^{-2})=\beta^2LF\neq 0$. 

This seems strange for two reasons. The torque is supposed to give the 
rate of change of the angular momentum. Even in the ${\cal O}'$ frame, 
each atom of the moving wedge is moving with constant velocity in the
$x$ direction only, so there is no rotation. Shouldn't that mean the
the angular momentum is constant\cite{Note2}?
Secondly, the angular momentum is part of a Minkowski tensor
$L^{\mu\nu}$, which we might expect to transform linearly, so as 
the spatial part is zero in the rest frame, and the $L^{0j}$ part is
at least constant in that frame, it again seems the ${\cal O}'$ ought
to see constant angular momentum components. 

One way some have chosen out of this conundrum is to argue that the
rate of change of  
angular momentum is given by the {\em total} torque, which needs to
include the torque due to internal forces. If we assume atoms in
different parts of the body apply equal and opposite forces on each other,
collinear in their mutual rest frame, this gives no additional torque
in that frame, but this pair of forces does produce a net torque
in frames with respect to which they are
moving. This is the approach taken by Nickerson and McAdory\cite{NMajp},
who take the angular momentum to mean the sum over atoms of their
individual $\vec L_i = \vec r_i\times \vec p_i$, where $\vec p_i = E_i
\vec u_i/c^2$, and $E_i$ is the total energy of the $i^\subrm{th}$
particle. 
Obviously this angular momentum is constant, and the total torque,
external plus internal, must sum to zero.
But this approach is unsatisfactory for several reasons. First, a
torque which requires knowledge of all the internal forces in a system
of particles is not very useful. Second, the particles in an interacting
system do not individually have a well-defined energy, and taking the 
system to include only the energy and momentum of each particle as if 
there were no interactions between them is inappropriate. Finally, the
treatment of a relativistic system as composed of particles with
action at a distance forces is dubious.

\section{Proper Treatment of Angular Momentum}

Our right-angle lever is stressed, and there is energy in that stress
which is
not included in the rest energy of each of the atoms. And while the 
momentum being injected into the system by the external forces totals
to zero, it is clearly being deposited in different locations from
where it is being extracted, so there is a flow of momentum, even in
the rest frame. A proper treatment needs to consider things locally.

We can resolve part of our conumdrum when we recognize
the angular momentum is not due only to the motion of the atoms of the lever.
In relativity global properties must be considered integrals over
local densities, and in particular the total momentum is the integral
over the lever of the momentum density. The momentum density is a
piece of the four-dimensional stress-energy tensor $T^{\mu\nu}$, which
has components 
\begin{eqnarray*}
  T^{00} &=& \hbox{ energy density}\\
  T^{j0} &=& c \times \hbox{ density of }(\vec P)_j, j=1, 2, 3\\
  T^{0j} &=& \frac 1c \times  \hbox{ flux of energy in the $j$ direction}\\
  T^{ij} &=& \hbox{ flux of $(\vec P)_i$, in the $j$ direction}.
\end{eqnarray*}
The 4-momentum $P^\mu =(E/c,\vec P)= \int F^{\mu0}(\vec x,t) d^3x$ of
an isolated
system is conserved because the stress-energy tensor is then a conserved
current, $\sum_\nu \partial_\nu T^{\mu\nu} = 0$. 

The angular momentum is given by integrating the moments of the
momentum density, 
\begin{eqnarray*}
L^{\mu\nu}(t) = \frac 1c \int d^3x\, {\cal M}^{\mu\nu 0}(\vec x,t)\\
\hbox{where}\quad {\cal M}^{\mu\nu\rho}(x) = x^\nu T^{\mu\rho}
- x^\mu T^{\nu\rho}.
\end{eqnarray*}
For an isolated system, conservation of angular momentum requires that
the angular momentum current is conserved, 
\begin{eqnarray*}
\partial_\rho {\cal M}^{\mu\nu\rho} = 0 &=&
\frac 1c \left(\delta^\nu_\rho T^{\mu\rho} + x^\nu \partial_\rho T^{\mu\rho}
-\delta^\mu_\rho T^{\nu\rho} - x^\mu \partial_\rho T^{\nu\rho}\right)\\
&=& \frac 1c \left(T^{\mu\nu} -T^{\nu\mu} \right),  
\end{eqnarray*}
where we have used that the energy-momentum current is itself
conserved, $\partial_\rho T^{\nu\rho} = 0$. 
Thus angular momentum conservation requires that the 
stress-energy tensor is symmetric, which we assume.

For observer ${\cal O}'$ there is an energy flow, 
because the $F_A$ force at the top is
doing work, injecting energy at a rate $F'_A v = Fv$ at the top, 
and the force $R'_x$ does negative work at the same rate, extracting
energy at the bottom. Thus while the two do not change the total
energy, they do produce a flow of energy. Across any $y = $ constant
surface between them there is a flow of energy $-Fv = cT^{0y}$. There
is no such net flow of energy in the $x$ direction due to the external
forces. Using the symmetry of the stress-energy tensor,
we see there is a momentum density $T^{y0}$ and 
$$ L_z = L^{xy} = \frac 1 c \int d^3x \left(x T^{y0} - yT^{x0}\right)$$
gets a contribution
$\Delta L^{xy} = - X FvL/c^2$, where $X$ is the mean $x$ coordinate of the
energy-flow down the arm. As this position is growing at a rate $v$, we have
$$ \frac {dL^{xy}}{dt'} = -\frac 1{c^2} Fv^2L = -\beta^2 LF,$$
consistent with the torque calculated from the forces in the ${\cal O}'$ frame.

What about spatial momentum? $F_x$ is injecting momentum in the x direction
in either reference frame, and it is being extracted by $R_x$, while 
$F_y$ and $R_y$ do the same in the $y$ direction. Thus there is a flow of 
momentum, but this does not contribute to $T^{j0}$.

So we have preserved
that torque is the rate of change of angular momentum. The necessity
of considering the angular momentum due to the energy flow, and thus 
resolving the paradox of nonzero torque with no rotation, was first
noticed by Laue\cite{LaueECDTR}

But what about tensor nature of $L^{\mu\nu}$? 
Our lever at rest had no $L^{ij}$ for spatial components,
but might it have had nonzero 
$$L^{0j}= \frac 1c
\int d^3x \left(x^j T^{0 0}(\vec r)-x^0 T^{j 0}(\vec r)\right)?$$
In the rest frame the first term is just $(Mc \vec R)^j$, where $M$ is the 
total mass (energy$/c^2$) and $\vec R$ the center of mass. The second
term vanishes 
as $T^{j 0}$ is the momentum density, which is zero. To ${\cal O}'$, however,
the first term is $\vecp R(t')  Mc\gamma$, while the second term is $-t'c P^j
=- t' Mc\gamma v$, so they add to a constant $Mc\gamma \vecp R(0)$. Thus 
in either frame $T^{j0}$ is constant, while $T^{ij}$ is constant in the rest
frame but not the moving frame. Thus the angular momenta we have
calculated are inconsistent with covariance.

\section{Covariance of Global Properties}

Now a well-defined physical property described by a covariant tensor
should not behave in this fashion. The problem is that we do not have
one well-defined property here. If we were talking about the momentum
and angular momentum of a point particle at a given space-time event,
the two observers would measure values consistent with covariance,
even if there were nonzero forces and torques acting on the particle,
because we would be comparing the values at the same event. But for an
extended object, when we describe a global property such as the total 
momentum or angular momentum, we are not talking about a single event,
but rather a sum or integral over different spatial points at some set
of times. Usually each observer uses the same time, according to his 
clocks, for all the points\cite{Note3}, and what is a single time
for one observer is not for another. The momentum and angular momentum
are integrals of a zero component of currents $J^\mu$, and if ${\cal O}$
examines the transformation back to his reference frame of what ${\cal O}'$
did to calculate his values, he would find ${\cal O}'$ had integrated
$J^\mu dS_\mu$ over a hypersurface $S$ of constant $t'$ and not one of 
constant $t$. If no external forces were acting, the current would be
conserved, $\partial_\mu J^\mu=0$, so Gauss' law would say the the
difference of these integrals is given by
a contribution from a surface 
connecting the two hyperplanes, which could be taken outside the body
and thus over a region with $J^\mu=0$. So both observers would agree.
But as there are external forces, this does not happen, $\partial_\mu
J^\mu\neq 0$ in the region between the two hypersurfaces,
so a contribution from the force at one end is included in the ${\cal O}'$
calculation, while the corresponding force (synchonous according to ${\cal O}$)
is not.

This issue is raised in Gamba \cite{Gamba} and Cavalleri and
Salgarelli\cite{CGnc}, who call our our approach definition (b) and synchronous
respectively, and contrast it to definition (a) or asynchronous
definitions of the total, which would have us integrate only over the 
hypersurface of times synchronous in the rest frame of the lever. Only
using that second definition of global quantities would one expect to get
covariance and constant angular momentum. Cavalleri {\it et.al.}\cite{CGSSajp} 
argue that failure to make this distinction led Nickerson and McAdory
\cite{NMajp} astray.

I would argue for the legitimacy of the synchronous definitions, as
long as one keeps in mind that using them, 
a time-dependent global property 
of an extended system is not a covariant object, even if the density
of that property is covariant.

In a recent paper\cite{Mansur} Masud Mansuripur has claimed that the 
Lorentz force law is incompatible with special relativity. He gives a 
simple example of a charge and magnetic dipole at rest in one frame, with
the dipole moment perpendicular to the line separating them. In the rest
frame there is no torque, but in a frame moving along their line of separation
the Lorentz force gives a net torque. It is claimed that this proves the
inadequacy of the Lorentz law, but our example shows this is not sufficient
to conclude that.

\section{An Alternate Example Clarifies Things}

To avoid the problem of computing the total angular momentum when external
forces are acting, consider the alternate scenario\cite{Kibble}
where the forces act
as described but only as an impulse at time $t=0$, simultaneously in
the ${\cal O}$ 
frame. That is, the force at A is $FT \delta(t) \hat e_x$ acting at
$\vec x=(0,L,0)$. We take the origin of ${\cal O}$ frame at O, and assume that
of ${\cal O}'$ coincides with O at $t=t'=0$. In the  ${\cal O}$ frame, the
forces deposit 4-momentum ($P^\mu = (E/c,\vec P)$) in the amounts of
$\Delta P^\mu=(0,FT,0,0)$ at A: $x^\mu=(0,0,L,0)$;
 $\Delta P^\mu=(0,-FT,-FT,0)$ at O: $x^\mu=(0,0,0,0)$; and 
$\Delta P^\mu=(0,0,FT,0)$ at B: $x^\mu=(0,L,0,0)$. 
In the  ${\cal O}$ frame these are all simultaneous and there is no 
change in total momentum or angular momentum.
In the ${\cal O}'$ frame, however, the momentum deposits and their space-time
locations are Lorentz transformed.
At A: $\Delta P^{\prime\,\mu}=(\beta\gamma FT, \gamma FT,0,0)$ 
at $x^{\prime\,\mu}=(0,0,L,0)$; 
at O: $\Delta P^{\prime\,\mu}=(-\beta\gamma FT, -\gamma FT,0,0)$ 
at $x^{\prime\,\mu}=(0,0,0,0)$; and
at B: $\Delta P^{\prime\,\mu}=(0, 0, FT,0)$ 
at $x^{\prime\,\mu}=(\beta\gamma L,\gamma L,0,0)$.

Before we discuss the angular momentum, let us comment on the position of 
what we usually call the center of mass, but which is really the center of
energy, 
$$\vecp R = \left.\int d^3r' \vecp r T^{\prime\,00}(\vecp r)\right/
\int d^3r' T^{\prime\,00}(\vecp r),$$
as $cT^{\prime\,00}$ is the energy density (for ${\cal O}'$). 
As the impulse at B occurs $\beta\gamma L/c$ after the 
other impulses, during that time interval, the wedge had momentum 
$P_y = -FT$, which meant during that interval the numerator in the center of
mass calculation was decreasing at a rate $c FT$ so it decreased by
$\beta\gamma LFT$. However the force at A injected energy $c\beta\gamma FT$ at
$y'=L$, which increased the numerator by $\beta\gamma LFT$ at $t'=0$, so there
is no net change at the end, and the $y$ position of the center of mass is 
unchanged.

We have already explained why the angular momentum can change even without 
any rotation. For ${\cal O}'$, the torque impulse at $t'=0$ from A is 
$\Delta L_z = -L(\gamma FT)$ at time $t'=0$, and $+(\gamma L)(FT)$ at 
$t'=\beta\gamma L/c$ from B, so overall there is no change. There are, however
instanteous changes at the ends, so during the interval 
$t'\in [0,\beta\gamma L/c]$,  there is a changed angular momentum. Initially,
this is due to the instantanous increase in $y_\tinyrm{cm}$, but during the 
interval this height is diminishing, and at the same time there is a flow of
negative $P_y$ injected at O towards B, where it will 
be\cite{Note4} cancelled by the 
impulse at B. This flow diminishes $L_z$ during the interval, so as to 
keep $L_z$ constant, until the torque impulse from B instantaneously restores
$L_z$ to its original value.

\begin{acknowledgments}
  The author wishes to thank Harold Zapolsky for exposing him to this
  issue and his reading and comments on the manuscript.
\end{acknowledgments}
\bibliographystyle{plain}

\end{document}